\documentclass[prl,twoside,twocolumn,showpacs,floatfix,nofootinbib,showkeys]{revtex4-1}

\usepackage{graphicx}
\usepackage{color}
\usepackage{epsfig}
\usepackage{amssymb}
\usepackage{amsmath}
\usepackage{bm}

\begin{document}

\bibliographystyle{apsrev}
\title{Period proliferation in periodic states in cyclically sheared jammed solids }
\author{Maxim O.~Lavrentovich}
\email{lavrentm@gmail.com}
\author{Andrea J. Liu}
\affiliation{Department of Physics and Astronomy, University of Pennsylvania, Philadelphia, Pennsylvania 19104, USA}
\author{Sidney R. Nagel}
\affiliation{Department of Physics, James Franck and Enrico Fermi Institutes, University of Chicago, Chicago, Illinois 60637, USA}
\begin{abstract}

Athermal disordered systems can exhibit a remarkable response to an applied oscillatory shear: after a relatively few shearing cycles, the system falls into a configuration that had already been visited in a previous cycle. After this point the system repeats its dynamics periodically despite undergoing many particle rearrangements during each cycle.  We study the behavior of orbits as we approach the jamming point in simulations of jammed particles subject to oscillatory shear at fixed pressure and zero temperature.  As the pressure is lowered, we find that it becomes more common for the system to find periodic states where it takes multiple cycles before returning to a previously visited state.   Thus, there is a proliferation of longer periods as the jamming point is approached.

\end{abstract}



\maketitle


Oscillatory sheared athermal particle packings or suspensions can fall into periodic ``absorbing states''~\cite{hinrichsen} in which the system returns to a configuration previously visited during the shearing process at the same point in the cycle. Once it returns to that configuration, the dynamics repeats itself indefinitely. At low densities in the absorbing state, the particles follow the flow without ever making contact with one another so that the system moves back and forth along a flat direction in the energy landscape \cite{pinegollub,cortepinegollub,nagelMemory1}, and particles return to their original positions after a single shear cycle: $T=1$.  As the strain amplitude $\gamma_t$ increases beyond some value $\gamma_t^*$, particles can no longer avoid each other and the system undergoes a dynamical ``absorbing state" transition from the absorbing phase to a phase in which the system continually visits new configurations. Models \cite{cortepinegollub,nagelMemory2,nagelMemory3,hexner,berthierDP} have linked this transition to variants of directed percolation~\cite{ramaswamy,berthierDP,berthierDP2}, which represents a broad class of non-equilibrium phase transitions~\cite{hinrichsen}. 

Athermal glasses such as Lennard-Jones glasses, by contrast, have an extensive entropy of energy minima that are not flat~\cite{stillinger,fractallandscape,bouchaud}. At very small strain amplitudes, they exhibit elastic behavior in which they explore different configurations within the same energy minimum. As $\gamma_t$ increases so that the system can explore more than one minimum, one might expect the system to meander indefinitely around a hopelessly intricate energy landscape as the system is driven in an oscillatory fashion.  Yet, remarkably, these systems can fall into absorbing states--they can find their way back to previously visited energy minima even as they undergo multiple particle rearrangements. Thus these systems explore many such minima~\cite{chaosyield,lookmandepin,royer,sastrymemory} over and over again.  Finally, when $\gamma_t$ is increased to $\gamma_t^*$, the system undergoes an absorbing state transition to a phase in which the system never returns to previously visited minima.

In this paper we investigate the fate of absorbing states in packings of jammed spheres that can be tuned to the jamming transition, where the system loses rigidity~\cite{jamreview,jamreviewVH}.  Far above this transition, absorbing states have a period of one cycle, so that the system returns to the same set of minima in each cycle.  With increasing strain amplitude $\gamma_t$, the number of minima explored before the system falls into an absorbing state increases so that there is a diverging time scale $\tau_a$ required for the system to fall into an absorbing state at $\gamma_t^*$. 

As the system approaches the jamming transition, the $\tau_a$ for the system to reach an absorbing state increases only weakly, by a factor of three over two orders of magnitude of pressure.  However, the {\emph{nature} of the absorbing state changes markedly--there is an increase in the number, $T$, of applied shear cycles between returns to a previous minimum. That is, there is a proliferation of higher-order periods, i.e., $T>1$, so that the same set of minima are explored in every set of $T$ consecutive shear cycles.  This result is consistent with the observation of multi-cycle periods in systems at densities just below jamming~\cite{belowjamming} as well as in frictional sphere packings very near jamming~\cite{royer}. Here we show that as the jamming transition is approached from the high-density side, the distribution of periods shifts systematically to higher $T$ while the value of $\gamma_t^*$ decreases far more weakly than the typical strain between rearrangements. As a result, the number of minima visited in each period increases quite rapidly, not only because the number of minima visited per cycle increases, but also because the number of cycles per period also increases. 

 \begin{figure}[h]
\includegraphics[width=3.41in]{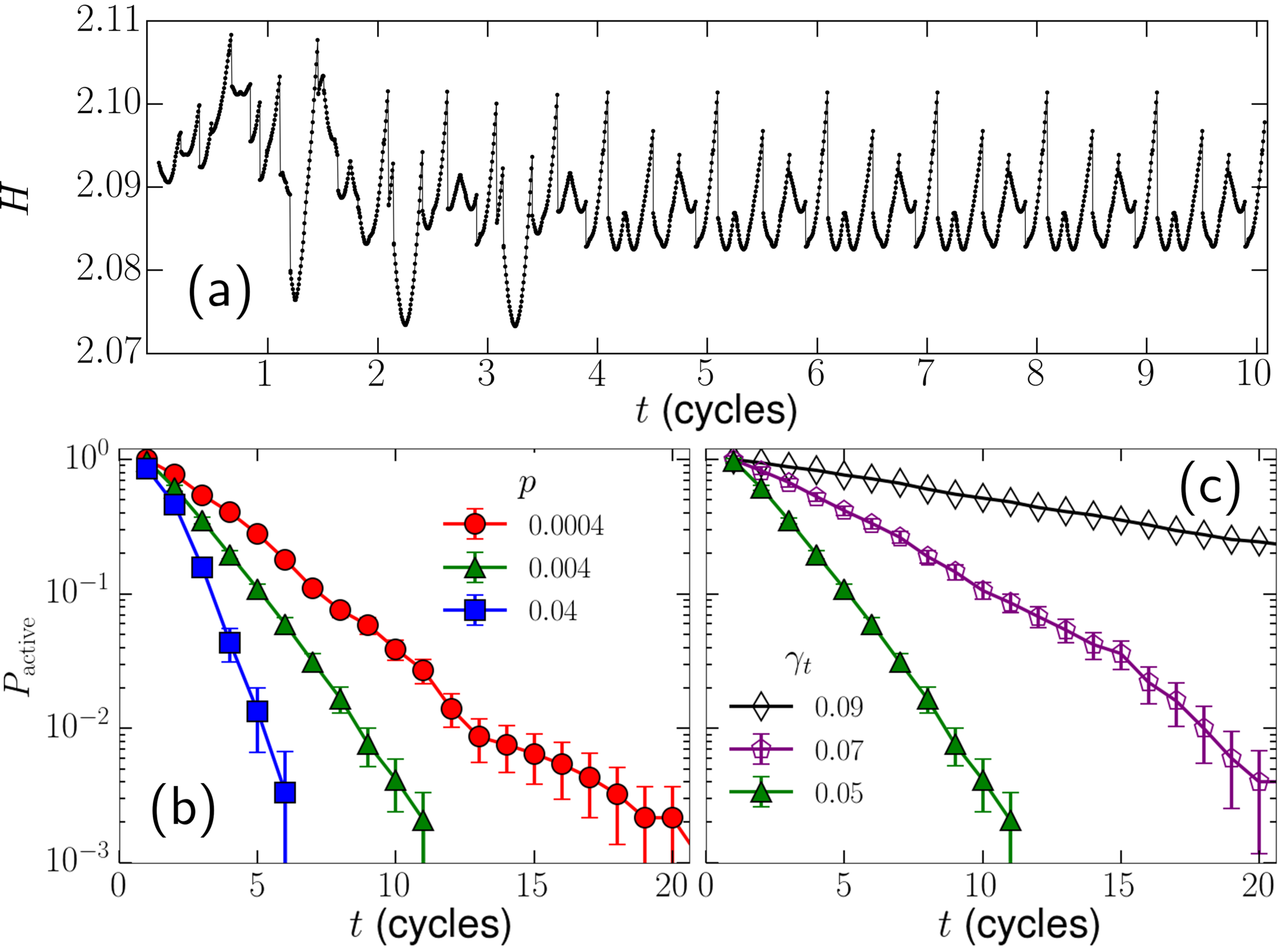}
\caption{\label{fig:shearIntro}  (a) The enthalpy $H$ as a function of time.  An oscillatory strain with amplitude $\gamma_t=0.15$ is applied quasi-statically to a system of $N=64$ particles.    (b, c) The probability $P_{\mathrm{active}}$ of a system (with $N=256$)  to still be ``active'' and not in a periodic cycle versus the number of applied shear cycles $t$, for a fixed  $\gamma_t=0.05$ in (b) and  for a fixed pressure $p\approx 0.004$ in (c).     The error bars for the $\gamma_t=0.09$ data are smaller than the symbol sizes.      }
\end{figure}

\textit{Simulations}.--- In our simulations, we study $N$ particles interacting with a  pair potential $U_{ij}$ acting between pairs of particles $i$ and $j$ (with radii $a_{i}$ and $a_j$) located at positions $\mathbf{r}_{i}$ and $\mathbf{r}_{j}$ ($i,j=1,2,\ldots,N$):
 \begin{equation}
 U_{ij}(r_{ij})= \frac{\epsilon}{\alpha}\left|1-\frac{r_{ij}}{A_{ij}}\right|^{-\alpha}\Theta(A_{ij}-r_{ij}),
\label{eq:pairpotential} \end{equation}
where $\epsilon$ is a characteristic energy, $r_{ij}=|\mathbf{r}_{i}-\mathbf{r}_{j}|$, $A_{ij}=a_i+a_j$, and $\Theta(x)$ is the Heaviside step function.   We use Hertzian interactions with $\alpha=5/2$ instead of the harmonic potential so that there is no discontinuity in the second derivative of the potential at the point of contact.  
   We study a mixture of particles with radii distributed uniformly between $a$ and $1.4a$ all with the same mass $m$. The units of length, mass, and energy are $a$, $m$, and $\epsilon$.    
The initial particle configurations were prepared by randomly placing the particle centers within the simulation box and then quenching to zero temperature using a fast inertial relaxation engine algorithm~\cite{FIRE} to relax the total energy.  The particle packing fraction was  adjusted to yield the desired pressure $p$, as in Ref.~\cite{CarlFS}, and the enthalpy (the appropriate thermodynamic potential for fixed pressure) was minimized at fixed  pressure.

We shear our configurations at fixed pressure, $p$, so that we can maintain the distance to the jamming transition during a shear cycle.  (Different particle configurations at the same packing fraction $\phi$ will generically have different values of the critical packing fraction  $\phi_c$  \cite{ohern1,ohern2}, so a constant volume ensemble  
does not keep the distance $|\phi-\phi_c|$ constant.)  To apply quasistatic shear at a constant pressure, we deform the system using a small strain step $\delta \gamma \sim 10^{-5}\mbox{-}10^{-6}$, and then minimize the enthalpy after each step.    The minimization process varies both the  volume $V$   and the relative particle coordinates $\mathbf{r}_{i}-\mathbf{r}_j$.  The volume is varied by scaling all particle positions by a  factor  $\beta_V$ ($\mathbf{r}_i \rightarrow \beta_V \mathbf{r}_i$),  which is related to a volume change $\delta V$ via $\beta_V= \sqrt{1+ \delta V/V}$.    The simulation box is taken to have Lees-Edwards periodic boundary conditions.  A similar procedure was used to generate packings at constant shear stress \cite{ningpotential}.    Note that by lowering $p$, we are able to approach the jamming point (since for our pair potential in Eq.~\ref{eq:pairpotential}, we expect that $p \propto |\phi-\phi_c|^{\alpha-1}=|\phi-\phi_c|^{3/2} $ \cite{ohern1}).  For each pressure and training amplitude, we average results over 300 to 1000 different initial configurations, until the quantities of interest have converged.

The strain is applied in both directions: the simulation box is sheared in one direction to a strain of $\gamma_t$,   then in reverse to $-\gamma_t$, and then back to zero again.   The resultant  enthalpy during the repeated application of such cycles is plotted in Fig.~\ref{fig:shearIntro}(a) for a single system. We see that  cyclic shear forces the system into a periodic state with $T=1$ after a four-cycle transient.  After the system becomes periodic, the enthalpy experiences various ``jumps,'' indicating the presence of rearrangements during which the system passes from one enthalpy minimum to another.   (We identify a rearrangement as any strain step during which the enthalpy changes by more than five times the typical enthalpy change.  As long as the strain steps are small enough, this procedure captures rearrangements and is insensitive to the particular threshold.) We are interested in the behavior of these systems as we increase the amplitude $\gamma_t$ and vary the pressure $p$.

To find the period $T$ of an absorbing state, we  track the enthalpy $H$ in stroboscopic snapshots taken at the end of each strain cycle.  As soon as a cycle has the  same $H$ as a previous cycle (i.e., the enthalpy difference is $10^{-5}$ times smaller than a typical enthalpy change during a shear cycle), then we count  the number of intervening cycles to find $T$.  We also verify that $H$ is the same every $T$ cycles for the remainder of the run. We track enthalpy rather than particle positions to assess whether the system is in an absorbing state, for two reasons. First, the method automatically excludes rattler particles, which do not contribute to the enthalpy but can complicate analysis in terms of particle positions because they do not have to return to the same position in each period even if the rest of the system has fallen into an absorbing state. Second, the method is less sensitive to numerical errors because small position fluctuations around the periodic states may lead to periodicity counts that depend sensitively on the threshold.  Such small fluctuations in the particle positions during a cycle were observed in simulations of frictional particle packings~\cite{royer}.

\begin{figure}
\includegraphics[width=3.41in]{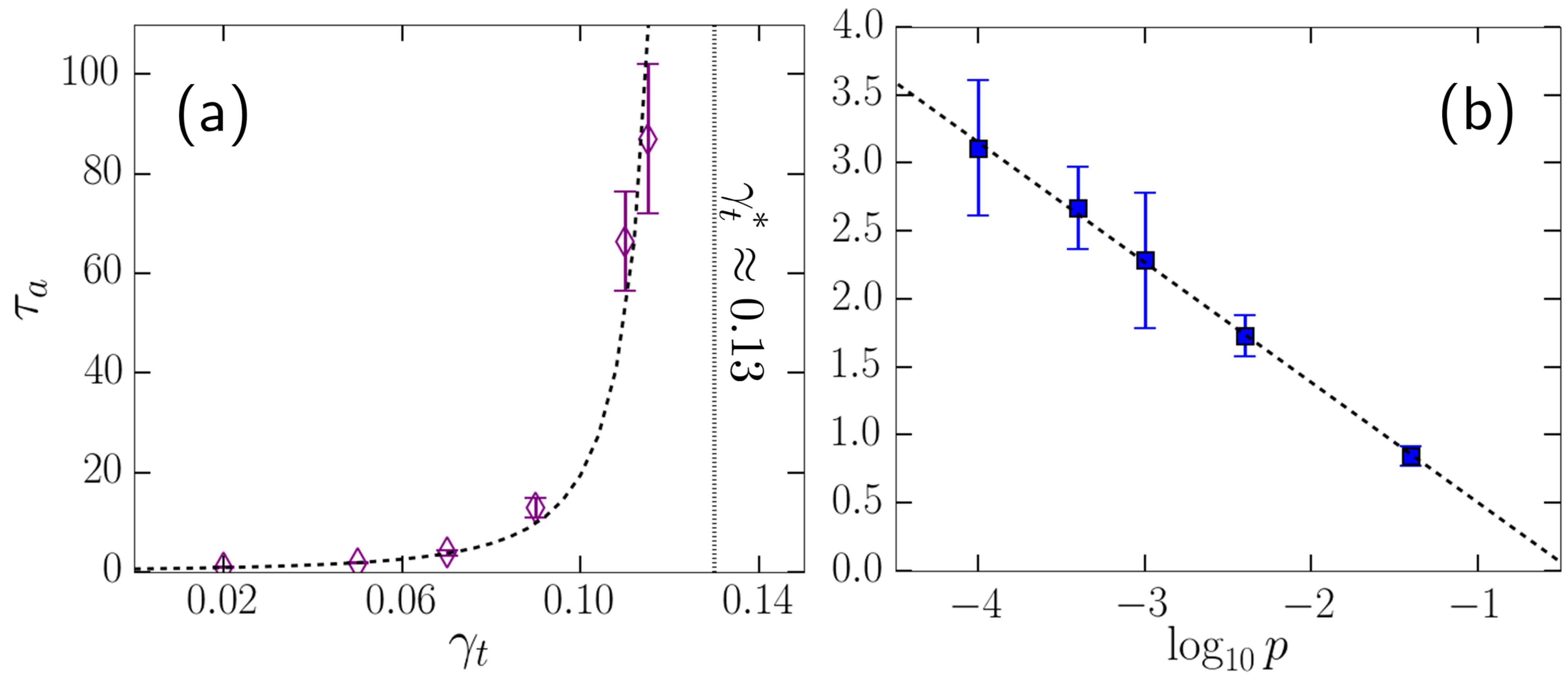}
\caption{\label{fig:chartime} Characteristic time $\tau_a$ to reach a periodic state, as a function of the training amplitude $\gamma_t$ (a) and pressure $p$ (b).   In (a), we see that $\tau_a$ appears to diverge at  a critical amplitude $\gamma_t^* \approx 0.13$ (dotted line). The dashed line shows a fit to $a (\gamma_t^*-\gamma_t)^{-\nu}$, with $\nu = 2.4 \pm 0.3$ and  $a =0.0043 $.   In (b), $\tau_a$ grows very slowly as $p$ decreases.  The dashed line shows a linear fit to the data  $\tau_a = a \log p+b$ with $a=-0.88$ and $b=-0.38$.}
\end{figure}

\textit{Results.}--- Figures~\ref{fig:shearIntro}(b,c) show that $P_{\mathrm{active}}$, the fraction of systems in the active (non-periodic) state, drops in an approximately exponential fashion to zero with the number of applied cycles.   This occurs both as we send the pressure $p$ to small values at fixed $\gamma_t$ [Fig.~\ref{fig:shearIntro}(b)] and as $\gamma_t$ is increased at fixed $p$ [Fig.~\ref{fig:shearIntro}(c)]. We may extract the characteristic time to approach a periodic state, $\tau_a$, from Figs.~\ref{fig:shearIntro}(b,c) by fitting the exponential decays ($\tau_a$ being the inverse of the decay constant), to yield the results in Fig.~\ref{fig:chartime}.  The errors are   estimated by fitting $\tau_a$ to different parts of the   $P_{\mathrm{active}}$ versus $t$ curves [see  Fig.~\ref{fig:shearIntro}(b,c)]. We see in Fig.~\ref{fig:chartime}(a) that the relaxation time $\tau_a$ appears to diverge as $\tau_a \sim (\gamma_t^*-\gamma_t)^{-\nu}$ with a critical exponent $\nu = 2.4 \pm 0.3$ at a finite critical training amplitude $\gamma_t^* \approx 0.13$. If $\gamma_t$ is larger than the critical value $\gamma_t^*$, the particles fail to return to a previously visited configuration before the end of the simulation run.    This is consistent with previous work on sheared packings above jamming~\cite{chaosyield}.  When the pressure $p$ is lowered, $\tau_a$ increases only slightly, consistent with $\tau_a \sim \ln (1/p)$, as shown in Fig.~\ref{fig:chartime}(b).

To investigate how system dynamics changes during the training protocol, we study the rearrangements in the first training cycle compared to those found in the eventual absorbing states.   The simplest way to quantify any differences is to measure the number, $N_r$, and magnitude $\Delta H$ of enthalpy drops  during a cycle.   We first compare the average change in enthalpy $\Delta H$  per rearrangement as a function of the training amplitude.    Figure~\ref{fig:scaling}(a) shows that magnitude of the enthalpy drops are suppressed by roughly a factor of 2 after the system has been trained into a periodic state.   We  also estimate  $\Delta H$ from systems under a continuous, steady-state shear at a fixed pressure $p\approx 0.004$.  In this case, we find an average enthalpy drop of $\Delta H_{\mathrm{con.}} \approx 4.6 \times 10^{-4}$.  This is larger than both the first training cycle and absorbing state enthalpy drops shown in Fig.~\ref{fig:scaling}(a):
For example, the first training cycle drops range from $0.97 \times 10^{-4}$ to $2.8 \times 10^{-4}$ in the data shown. We expect that the first training cycle drops will converge to $\Delta H_{\mathrm{con.}}$ as we increase the training amplitude, but our amplitudes are less than $0.1$ and far from this convergence.

\begin{figure}
\includegraphics[width=3.45in]{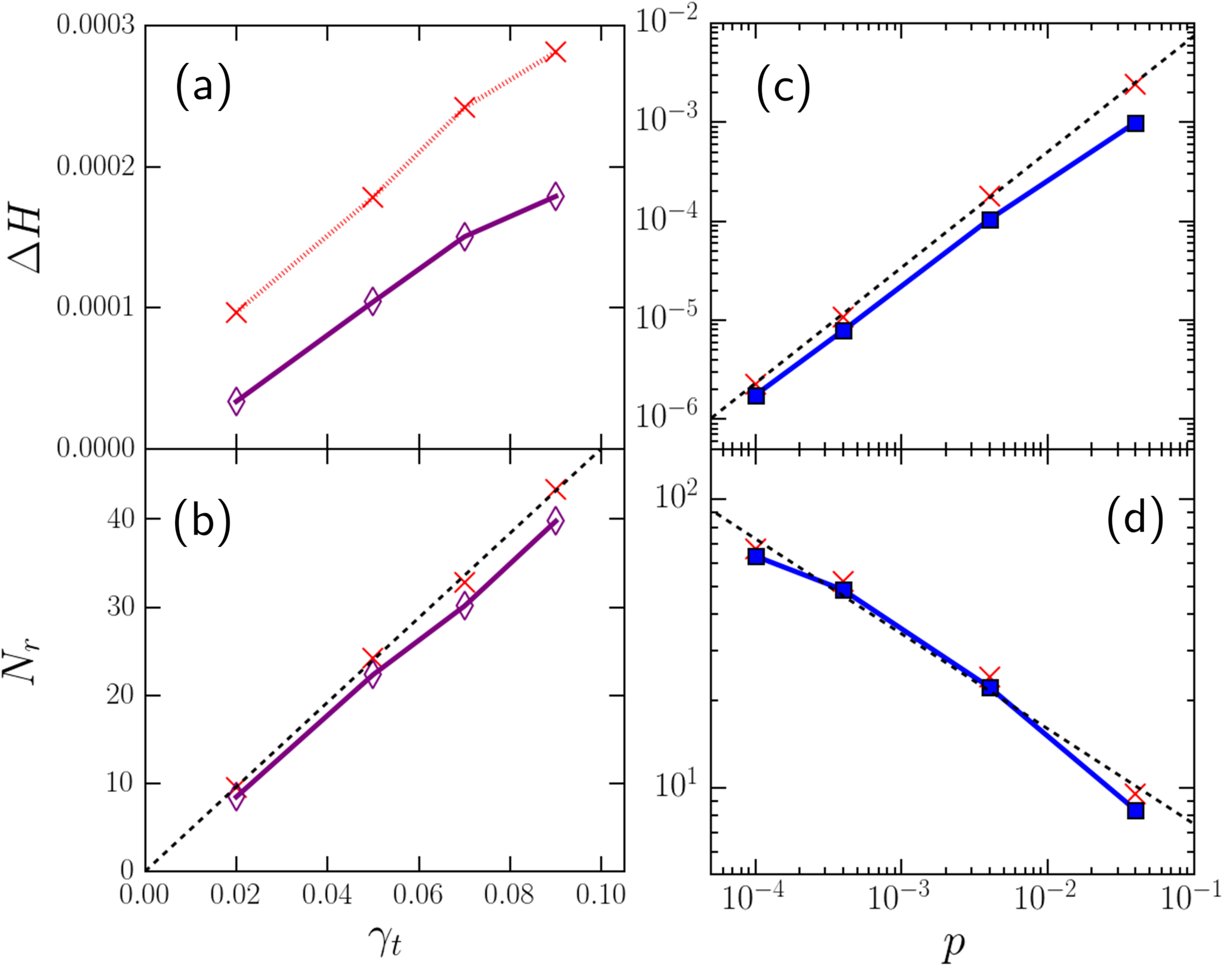}
\caption{\label{fig:scaling}  Comparison for a system with $N=256$  of the first training cycle (red Xs) with the eventual periodic cycle (purple diamonds and blue squares) into which it settles, for fixed pressure $p\approx 0.004$ and varying training amplitude $\gamma_t$ in (a,b) and for a fixed  $\gamma_t = 0.05$ (and varying $p$) in (c,d).    We plot  the average enthalpy drops $\Delta H$ during particle rearrangements in (a,c) and the number of rearrangements $N_r$ in  (b,d).  In (b), the dashed line shows a linear fit through the first cycle data: $N_r=480 \gamma_t$.    In (c, d), the dashed lines show power-law fits to the first cycle data, with  $\Delta H = 0.11 p^{1.17}$ and  $N_r =3.5 p^{-0.33}$, respectively. The errors in the data points are smaller than the symbol sizes.          }
\end{figure}

Figure~\ref{fig:scaling}(b) shows that the average number of rearrangements per cycle, $N_r$, does not vary appreciably between the first cycle and the periodic state.  Also,  $N_r$ scales linearly with the training amplitude $\gamma_t$.  Therefore, we may think about these rearrangements occurring at a constant rate, with an average yield strain separating successive rearrangements
\begin{equation}
\gamma_y \approx 4 \gamma_t/N_r , \label{eq:yieldstrain}
\end{equation}
(since $4 \gamma_t$ is the total strain during a cycle).  Figure~\ref{fig:scaling}(b) shows a linear fit  $\gamma_y \approx 8.3 \times 10^{-3}$ for the first cycles in systems with  $p \approx 0.004$  (dashed line). Our strain step size $\delta \gamma \sim 10^{-5}$ was chosen to be much smaller than this value. 

   Figure~\ref{fig:scaling}(c) shows that the average enthalpy drop during a rearrangement in the first cycle increases with the pressure:  $\Delta H \sim p^{1.17 \pm0.06}$.
As in Fig.~\ref{fig:scaling}(a), we find  a relatively  small suppression of $\Delta H$ for the periodic states compared to the first training cycle, with the biggest differences (suppression by about a factor of 2) occurring at higher pressures. 
Figure~\ref{fig:scaling}(d) shows that at fixed amplitude of strain, $N_r \sim p^{-0.33 \pm 0.09}$.  This is consistent with an argument based on using the scaling properties near the jamming transition.  The static shear modulus $G$ scales with the distance to the jamming point according to $G \sim (\Delta \phi)^{\alpha-3/2} \sim p^{2/3} $ for Hertzian interactions, $\alpha= 5/2$~\cite{ohern1}.  Next, the yield stress $\sigma_y$ that induces a rearrangement should be given by  $\sigma_y \propto (\Delta \phi)^{\alpha-1} \propto  p$~\cite{YieldStressScaling} in the quasistatic limit.  Therefore, the yield strain satisfies $ \gamma_y \sim \sigma_y/G \sim p^{1/3}$.  
In a cycle, then, we would expect the number of rearrangements to scale according to Eq.~\ref{eq:yieldstrain}: $N_r \sim 1/ \gamma_y \sim p^{-1/3}$, in reasonable agreement with our results.

In summary, Fig.~\ref{fig:scaling} shows that the absorbing states explore many different minima in the landscape, and that enthalpy drops during transitions between minima are not appreciably smaller than those in the initial training cycle.  Similar behavior was recently observed in finite-temperature simulations of Lennard-Jones glasses~\cite{yieldsastry}, where the avalanche size statistics provide no signal of the absorbing-state transition. Together, these results show that even multiple, quite extended particle rearrangements with large enthalpy drops are precisely cancelled out so that the system returns to the same configuration at the end of each period.  The presence of these delicate balances likely leads to sensitivity to perturbations in particle positions as observed in constant-volume simulations \cite{chaosyield}.   Nevertheless, our results show that the \emph{statistics} of the absorbing states depend systematically on pressure and training amplitude.

\begin{figure}[h]
\includegraphics[width=2.8in]{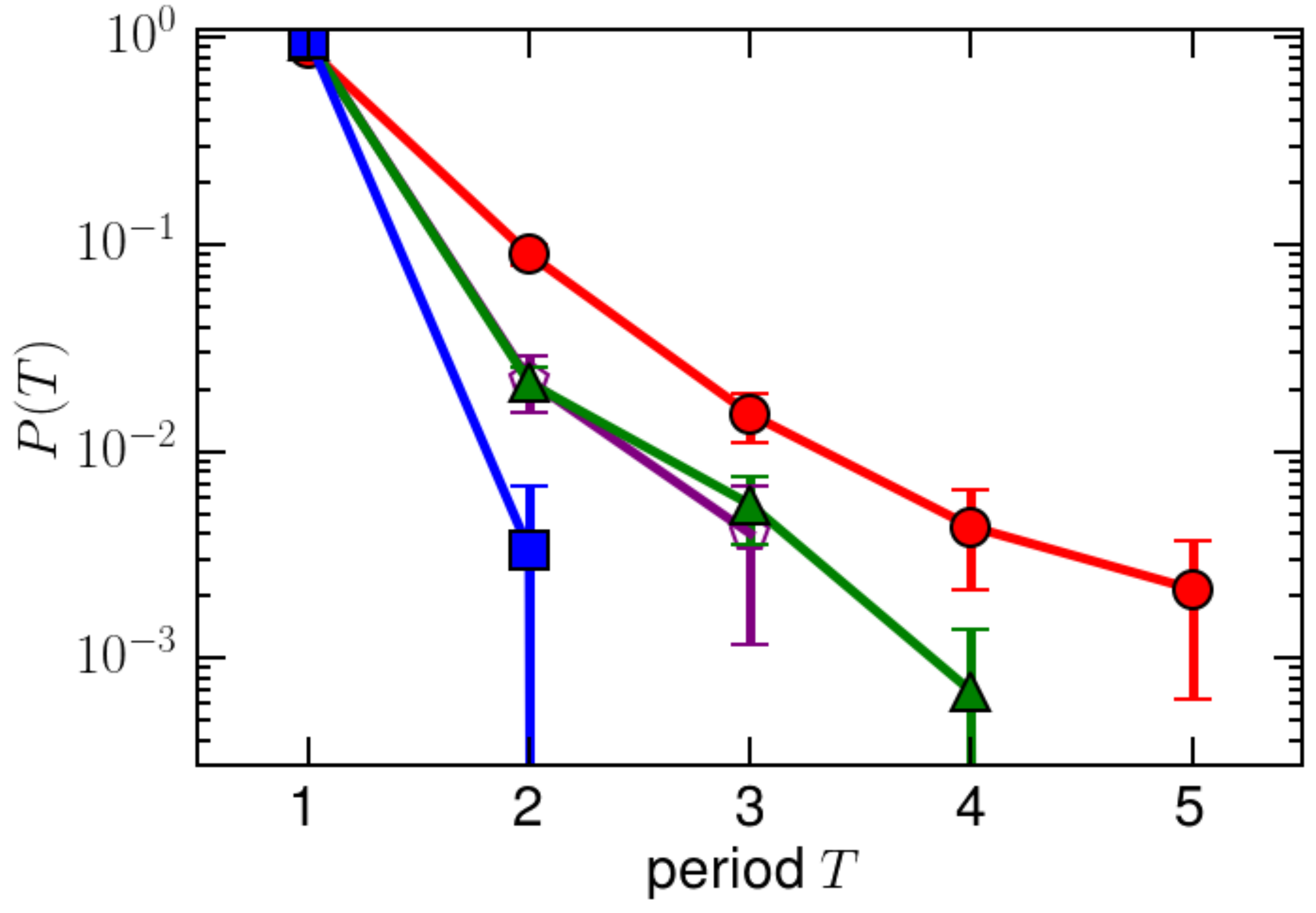}
\caption{\label{fig:HiLowP}   The fraction of  $N=256$-particle systems, $P(T)$ that have settled into a periodic cycle with period $T$, at training amplitude $\gamma_t=0.05$ and pressures $p=0.04$ (blue squares), $p=0.004$ (green triangles) and $p=0.0004$ (red circles).  Lines are to guide the eye. We also include results at $p \approx 0.004$ at a higher training amplitude of $\gamma_t=0.07$  (open purple pentagons). The fraction of systems with multi-cycle periods increases with decreasing $p$. }
\end{figure}

By studying the periodicity of the absorbing states, we find a qualitative change in the dynamics that arises as the system approaches the jamming transition ($p \rightarrow 0$).  Figure~\ref{fig:HiLowP} shows that as $p$ is lowered, the system settles into periods with higher numbers of cycles $T$.  That is, the system must undergo multiple cycles of the applied shear before it returns to the same configuration. 

It has been suggested that period proliferation occurs with increasing $\gamma_t$~\cite{chaosyield}. We do not observe this; when the training amplitude is raised at a fixed pressure, the periodicity of the absorbing states remains roughly the same, as shown in Fig.~\ref{fig:HiLowP}, where two distributions for $\gamma_t=0.05$ and $\gamma_t=0.07$ are shown (green triangles and purple pentagons) for a fixed pressure: $p \approx 0.004$.  The two distributions are the same within the simulation error, suggesting that it is primarily the pressure that controls the periodicity increase, not training amplitude. Therefore, we have two ways in which we may lose the simplest mode of periodicity, $T=1$: There can be a proliferation of higher $T$ periodicities as $p \rightarrow 0$ and there can be a diverging time $\tau_a$ to reach the absorbing state as $|\gamma_t -\gamma_t^* | \rightarrow 0$.

Multi-cycle periods have been observed before in simulations of systems near the jamming transition, either below jamming~\cite{belowjamming} or near jamming in packings of frictional particles~\cite{royer}. Our results show that the periodicity of absorbing states can be tuned systematically by varying the pressure, or equivalently, the distance from the jamming transition. 

Note that at $p=4 \times 10^{-4}$, Fig.~\ref{fig:scaling}(d) shows that there are approximately 50 rearrangements per cycle, when averaged over all absorbing states studied. Some of these absorbing states have $T \geq 5$, so that the system can visit 250-300 minima before repeating itself. 

\textit{ Discussion.}---  In summary, we have characterized changes in the properties of absorbing states on approach to the jamming point. Remarkably, the absorbing states are similar in their statistics to the states visited during an initial shear cycle: absorbing states can have large rearrangements which must precisely cancel to yield a periodic state.  Lowering the pressure does not appear to lower the probability of finding an absorbing state. Even the time needed for the system to fall into an absorbing state increases only gradually with decreasing pressure. Instead, we find that the absorbing states become more complicated. There is a proliferation of multi-cycle, $T>1$, absorbing states as the system approaches the jamming transition.  Multi-cycle absorbing states are also observed in systems that approach the jamming transition from the low-density side~\cite{belowjamming}.  The possibility that the jamming transition might correspond to the point of maximum period proliferation is intriguing.  The nature of the absorbing state gives us indirect information about the energy landscape; perhaps future studies will make this connection explicit.

\begin{acknowledgments}

We thank Carl P. Goodrich, Daniel Hexner, Sri Sastry and Daniel M. Sussman for discussions and help with the simulations.  We acknowledge support from the  UPENN MRSEC under Award No. NSF-DMR-1120901, NSF Grant DMR-
1262047 (MOL), NSF Grant DMR-1404841 (SRN),  and computational support from the Simons Foundation for the collaboration ``Cracking the Glass Problem'' (454945 to AJL and 348125 to SRN).  In addition, this research was supported by the US Department of Energy, Office of Basic Energy Sciences, Division of Materials Sciences and Engineering under Awards DE-FG02-05ER46199 (AJL). Additional computational support was provided by the Walnut HPC Cluster at the University of Pennsylvania. 
\end{acknowledgments}

\bibliographystyle{apsrev}

\end{document}